\documentclass[a4paper,11pt]{article}
\pdfoutput=1 

\usepackage{jinstpub} 
\usepackage{isotope}
\usepackage{lineno}

		\title{Fluorescence emission of the JUNO liquid scintillator}
		
		\author[b,1]{M.\,Beretta\note{Corresponding author.}}
		\author[b]{F. Houria}
		\author[b]{F.\,Ferraro}
		\author[b]{D. Basilico}
		\author[b]{A. Brigatti}
		\author[b]{B. Caccianiga}
		\author[b]{A. Caslini}
		\author[b]{C. Landini}
		\author[b]{P. Lombardi}
		\author[b]{L. Pelicci}
		\author[b]{E. Percalli}
		\author[b]{G. Ranucci}
		\author[b]{A.C. Re}
		\author[f]{C. Clementi}
		\author[f]{F. Ortica}
		\author[f]{A. Romani}
		\author[b]{V. Antonelli}
		\author[b]{M.G. Giammarchi}
		\author[b]{L. Miramonti}
		\author[b]{P. Saggese}
		\author[b]{M.D.C. Torri}
		\author[a]{S. Aiello}
		\author[a]{G. Andronico}
		\author[e]{A. Barresi}
		\author[c]{A. Bergnoli}
		\author[e]{M. Borghesi}
		\author[c]{R. Brugnera}
		\author[a]{R. Bruno}
		\author[d]{A. Budano}
		\author[j]{A. Cammi}
		\author[c]{V. Cerrone}
		\author[a]{R. Caruso}
		\author[e]{D. Chiesa}
		\author[e]{C. Coletta}
		\author[c]{L.V. D'Auria}
		\author[c]{S. Dusini}
		\author[d]{A. Fabbri}
		\author[g]{G. Felici}
		\author[c]{A. Garfagnini}
		\author[a]{N. Giudice}
		\author[c]{A. Gavrikov}
		\author[c]{M. Grassi}
		\author[a]{N. Guardone}
		\author[c]{L. Lastrucci}
		\author[c]{I. Lippi}
		\author[j]{L. Loi}
		\author[a]{C. Lombardo}
		\author[h,i]{F. Mantovani}
		\author[d]{S.M. Mari}
		\author[g]{A. Martini}
		\author[h,i]{M. Montuschi}
		\author[e]{M. Nastasi}
		\author[d]{D. Orestano}
		\author[g]{A. Paoloni}
		\author[d]{F. Petrucci}
		\author[e]{E. Previtali}
		\author[h,i]{B. Ricci}
		\author[a]{G. Sava}
		\author[c]{A. Serafini}
		\author[c]{C. Sirignano}
		\author[e]{M. Sisti}
		\author[c]{L. Stanco}
		\author[d]{E. Stanescu Farilla}
		\author[h,i]{V. Strati}
		\author[a]{C. Tuve'}
		\author[d]{C. Venettacci}
		\author[a]{G. Verde}
		\author[g]{L. Votano}
		
		\affiliation[b]{INFN, Sezione di Milano e Università degli Studi di Milano, Dipartimento di Fisica, Italy}
		\affiliation[a]{INFN, Sezione di Catania e Università di Catania, Dipartimento di Fisica e Astronomia, Italy}
		\affiliation[c]{INFN, sezione di Padova e Università di Padova, Dipartimento di Fisica e Astronomia, Italy}
		\affiliation[d]{INFN, sezione di Roma Tre e Università degli Studi Roma Tre, Dipartimento di Matematica e Fisica, Italy}
		\affiliation[e]{INFN, Sezione di Milano Bicocca e Dipartimento di Fisica Università di Milano Bicocca, Italy}
		\affiliation[f]{INFN, Sezione di Perugia e Università degli Studi di Perugia, Dipartimento di Chimica, Biologia e Biotecnologie, Italy}
		\affiliation[g]{Laboratori Nazionali dell’INFN di Frascati, Italy}
		\affiliation[h]{INFN, Sezione di Ferrara, Italy}
		\affiliation[i]{Università degli Studi di Ferrara, Dipartimento di Fisica e Scienze della Terra, Italy}
		\affiliation[j]{INFN, Sezione di Milano Bicocca e Dipartimento di Energia, Politecnico di Milano, Italy}

\emailAdd{marco.beretta@mi.infn.it}

\abstract{JUNO is a huge neutrino detector that will use 20 kton of organic liquid scintillator as its detection medium. The scintillator is a mixture of linear alkyl benzene (LAB), 2.5 g/L of 2,5-diphenyloxazole (PPO) and 3 mg/L of 1,4-Bis(2-methylstyryl)benzene (Bis-MSB). The main goal of JUNO is to determine the Neutrino Mass Ordering \cite{JUNO, solari, SN}. In order to achieve this purpose, good energy and position reconstruction is required, hence a complete understanding of the optical characteristics of the liquid scintillator is mandatory. In this paper we present the first measurements on the JUNO scintillator emission spectrum, absorption length and fluorescence time distribution performed respectively with a spectrofluorimeter, a spectrophotometer and a custom made setup of the liquid scintillator produced by the JUNO purification plants. }

\keywords{Liquid detectors, Scintillators, scintillation and light emission processes, Large detector systems for particle and astroparticle physics, Neutrino detectors}

\begin{document}
\maketitle
\flushbottom
\section{Introduction}

Over the past few decades, liquid scintillator based detectors have been widely employed in particle and astroparticle physics for rare event searches, aiming to detect dark matter \cite{Kumar:2015nj} and neutrinos from natural \cite{BOREXINO:2022abl} and artificial sources \cite{DayaBay:2018yms}. These experiments have been fundamental for the determination of the neutrino oscillation parameters \cite{DayaBay:2018yms,SNO_Oscillations} and solar neutrino spectroscopy \cite{BOREXINO:2022abl,Borexino:2017rsf,SNO:2003bmh}.



\noindent Liquid scintillator detectors are usually instrumented with photomultiplier tubes (PMTs) to detect the scintillation light produced as a charged particle passes throughout the medium. The optical characterization of the detection medium is crucial  to properly reconstruct the physics events and to discriminate the signal from internal or the muon-induced radioactive backgrounds. In particular, the determination of fluorescence emission time is mandatory to accurately reconstruct the interaction position.
Also, the time profile of the fluorescence light emitted by a scintillator depends on the particle type: this characteristic is commonly used in the Pulse Shape Discrimination (PSD) technique to perform particle discrimination \cite{BX_PSD}. 
In this work we present the measurements for the optical characterization of the liquid scintillator that will be used in the JUNO experiment. In particular we study a sample deriving from the commissioning of its purification plants \cite{LANDINI}.

\noindent We will focus on the emission spectrum (section \ref{sec:emission_spectra}), on the absorbance (section \ref{sec:absorbance}) and on the characteristic fluorescence time profile (section \ref{sec:fluorescence}). Other properties, like the refractive index of the JUNO liquid scintillator and the LAB absorption length at wavelengths around 430 nm, have been published in other works \cite{rewind, LAB_abs}. 

\section{Composition and properties of linear alkylbenzene based liquid scintillators}

When a charged particle deposits energy in an organic scintillator, it causes the excitation of electrons in the $\pi$-bonds of the solvent molecules \cite{KNOLL}.
Electrons can populate different excited states, that quickly decay non-radiatively to the first excited singlet state with a typical lifetime of the order of $\sim$ps; the latter then decays to the ground state, emitting fluorescence light with typical lifetimes of the order of few ns (fast component). As an alternative, electrons can also populate the first excited triplet state and then transit to the first excited singlet state, because of thermal excitation, and generate delayed fluorescence with a time scale of the order of $\sim$ hundreds of ns (slow component). Both the fast and the slow components generate the fluorescence signal we are interested in.
Note that only $\sim5\%$ of the particle energy is converted into light.




\noindent To increase the photon emission, a second component, usually called  ``primary wavelength shifter", is added to the solvent in small mass concentrations ($\sim$g/L). The solvent transfers the deposited energy to the primary wavelength shifter, mainly following non-radiative processes, that subsequently emits fluorescence light at longer wavelengths above the absorption spectrum of the solvent.
Usually, large mass detectors are equipped with photomultiplier tubes (PMTs) to detect the scintillation photons. To better match the efficiency curve (quantum efficiency Q.E.) of the PMTs, a third component (the so-called ``secondary wavelength shifter") can be added in even smaller quantities ($\sim$mg/L). 

\noindent As a result of the excitation and de-excitation of the components in the mixture, the time distribution of scintillation light can be effectively modelled by a linear combination of a few exponential contributions, each one associated to a characteristic decay time.

\noindent Over the last decades, linear alkylbenzene has been established as one of the best available solvents for a liquid scintillator detector, since it is not toxic, not flammable and it is available in large quantity at relatively low cost. 
These features are crucial in large detectors for the search of rare events. For this reason LAB was chosen for the Daya Bay \cite{DB}, RENO \cite{RENO} and SNO+ \cite{SNO+} experiments and it will also be employed by new generation of experiments such as JUNO \cite{JUNO} and SABRE \cite{SABRE}.
2,5-diphenyloxazole (PPO) was chosen as the primary wavelength shifter in these experiments with a concentration of 2.5 g/L, 2 g/L and 3 g/L respectively. Moreover JUNO and SABRE will also use 1,4-Bis(2-methylstyryl)benzene (Bis-MSB) with concentrations of 3 mg/L and 15 mg/L respectively as secondary wavelength shifter \cite{OptModel}. Bis-MSB is planned to be used in the Tellurium phase of the SNO+ detector for the $0\nu \beta \beta$ decay search \cite{SNO+}. A summary of the liquid scintillator mixtures employed in these experiments is reported in table \ref{tab:mixtures}.
\begin{table}[tbh]
	\centering
	\caption{Composition of a selection of LAB-based scintillators used in past, present and future experiments.}
	\begin{tabular}{lcc}
		\hline
		Mixture     &   PPO (g/L)   &   Bis-MSB (mg/L)  \\
		\hline
		Daya Bay (+ Gd)         &   3.0           &   15              \\
		RENO        & 3.0           & 30 \\
		JUNO        &   2.5         &   3               \\
		SNO+         &   2.0           &   -               \\
		SNO+ Te-phase & 2.0           & 15                \\
		SABRE       & 3.5           & 15                \\
	\end{tabular}
	\label{tab:mixtures}
\end{table}
\newline
\noindent The amount of flours in the scintillator mixture determines its optical properties, including the time distribution of fluorescence light, the light yield and the emission and absorption spectra.


\section{Emission spectrum}\label{sec:emission_spectra}
In a liquid scintillator based experiment, the emission spectrum has a decisive impact on the achievable energy resolution. For LAB-based scintillators, the emission spectrum depends on the chemical composition of the solvent. LAB ($\mathrm{C_6H_5C_nH_{2n+1}}$ with $10\leq \mathrm{n} \leq 16$) is generally a mixture of different chemical components. Since the proportion of this mixture may change depending on the producer, the emission spectrum may be significantly different depending on the procurement source. All our measurements are referred to the LAB bought specifically for the JUNO experiment and processed in the JUNO purification plants \cite{LANDINI}. 
The spectral shape of the emission spectrum depends also on the proportion of PPO and Bis-MSB in the mixture. Both components are used to optimize the coupling between the PMTs and the liquid scintillator emission, in order to have the highest light collection possible.

\subsection{Setup}

The emission spectra of the scintillator mixtures were measured using a Spex Fluorolog-2 1680/1 spectrofluorimeter, controlled by the Spex Datamax spectroscopy software, similarly to what was described in \cite{PAOLO}.
The excitation light at 260 nm was produced by a Xenon discharge lamp with a monochromator. The resulting emission light was then sampled with a second monochromator. The monochromators were set to have a bandwidth of 6 nm and 3 nm respectively.

\noindent Measurements have been performed with a right angle geometry (at 90° angle) with respect to the direction of the incident light. This allows to look at the production of light in the inner part of the samples allowing to measure both the effects of PPO and Bis-MSB.

\noindent The quantum efficiency of the PMT which detects the light is taken into account thanks to a measurement with a Rhodamine calibration standard in the spectral region of interest \cite{PAOLO}.
Each point in the spectrum is normalized to the excitation light intensity, thanks to a beam splitter and a reference detector.

\begin{figure}
	\centering
	\includegraphics[width=0.4\linewidth]{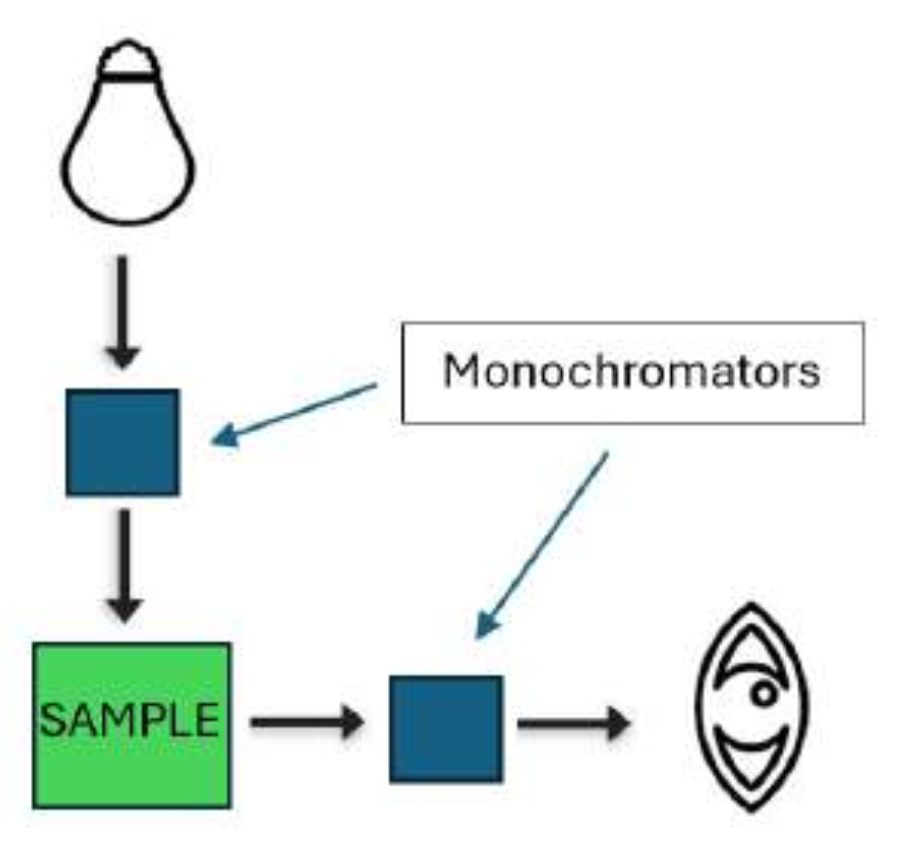}
	\caption{Schematic view of the spectrofluorimeter. The light emitted by the Xenon lamp is passed to a monochromator before hitting the sample. Then the light is collected in a 90° geometry and collected by a photomultiplier tube.}
	\label{fig: Fluorimetro}
\end{figure}

\subsection{Measurements}

\begin{figure}[h]
	\centering
	\includegraphics[width=0.6\columnwidth]{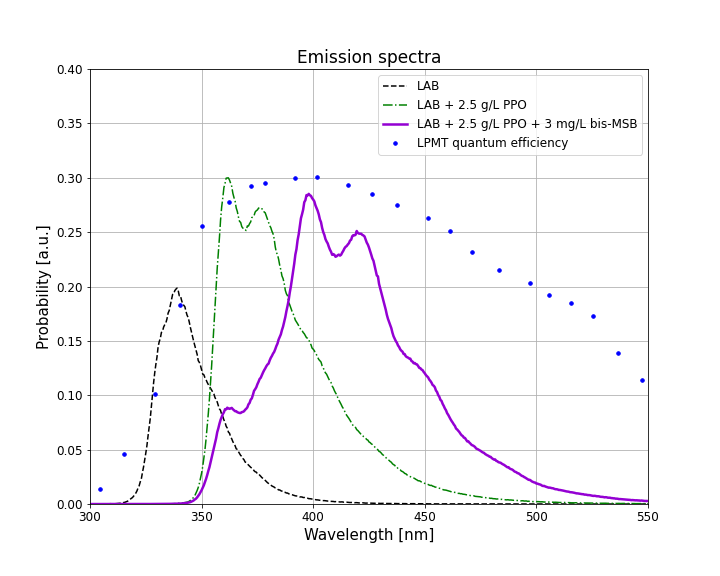}
	\caption{Emission spectra of the scintillator mixture (in violet solid line) and of its components (LAB in dashed black line and LAB + 2.5 g/L in dash-dotted green line). The plot also shows the JUNO large NNVT-PMT (LPMT) quantum efficiency in blue dots, reported in \cite{JUNO_prediction}. } 

\label{fig:EmissionSpectra}
\end{figure}

We measured the emission spectrum of the JUNO liquid scintillator between 250 nm and 600 nm in a 1 cm-path length cuvette in a right angle geometry, as shown in figure \ref{fig: Fluorimetro} (figure \ref{fig:EmissionSpectra} shows a zoom between 300 nm and 550 nm). One can see how the emission spectrum changes when a component is added to the LAB. The violet, green and black lines show the emission spectra of LAB + 2.5 g/L PPO + 3 mg/L Bis-MSB (the recipe of the JUNO liquid scintillator), LAB + 2.5 g/L PPO and pure LAB. We note that the  spectrum reported here  has been obtained in a small geometry setup ($\sim$ centimenters). In JUNO, due to its large dimensions ($\sim$ several tens of meters) the situation will be significantly different: in particular, the lower part of the spectrum will be absorbed and re-emitted above 400 nm. This would make the peaks at 400 and 420 more dominant and the total emission more similar to the bis-MSB one \cite{OptModel}.

\noindent The emission spectrum for the full mixture (violet solid line) peaks at 400 nm and 420 nm. The two smaller peaks at about 360 nm and 375 nm are residual of the PPO emission and in fact they dominate the LAB + 2.5 g/L PPO emission (green dashed line).
It is important to notice that the emission of the JUNO mixture matches well the quantum efficiency of the 20" NNVT-PMTs that are employed in the detector (represented by the blue dots in figure \ref{fig:EmissionSpectra}).\\

\section{Absorbance and absorption length}\label{sec:absorbance}
Another optical property which plays a crucial role in large detectors, like JUNO, is the absorbance $A$ of the liquid scintillator, which is related to the absorption length $L_A$ as follows:  
\begin{equation}
A = - \log_{10}{\frac{I}{I_0}}  = - \log_{10}{e^{-\frac{x}{L_A}}}
\label{eq:absorbance}
\end{equation}
where $I_0$ is the incident light intensity, $I$ is the light intensity after the propagation and $x$ is the optical path length. 
A large absorbance in the region of the emission peak of the scintillator implies a significant loss of photons, which in turn has a significant impact on the energy resolution of the detector.
The absorbance of the scintillator mixtures has been measured in a range between 200 nm and 700 nm, using a Jasco Dual Ray V-760 Spectrophotometer provided with a LSE-701 Single Position Long Path Length Cell Holder (maximum length of the holder: 100 mm).
The measurements have been conducted using a blank sample of hexane as a reference and a 10 cm long cuvette. Hexane has an high transparency and a refractive index similar to the scintillator. This helps minimizing propagation effects (like reflection and refraction) which can bias the absorbance measurements. 

\begin{figure}[h]
\centering
\includegraphics[width=0.7\columnwidth]{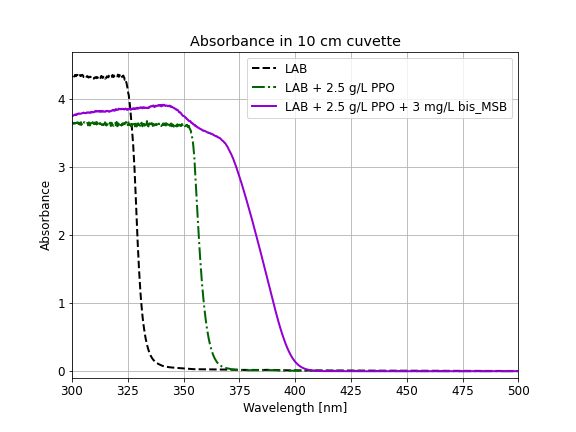}
\caption{Absorbance of the JUNO liquid scintillator mixture (in solid violet line) compared to the LAB (in dashed black line) and LAB + 2.5 g/L PPO (in dash-dotted green line) absorbances in the region of interest.}
\label{fig:AbsorptionSpectra}
\end{figure}

\noindent We studied the impact of progressively adding components to the mixture, like PPO and Bis-MSB. As shown in figure \ref{fig:AbsorptionSpectra}, the JUNO mixture has a significant absorbance up to 400 nm compared to the raw LAB. This is due to the wavelength shifters, PPO and Bis-MSB, which by design absorb photons in the region of the solvent emission, in order to re-emit them at higher wavelengths.

\begin{figure}[h]
\centering
\includegraphics[width=\columnwidth]{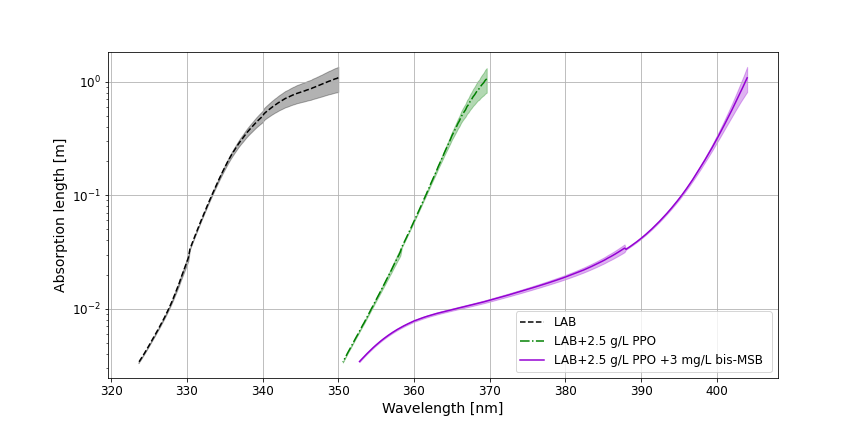}
\caption{Error-weighted mean of the attenuation length with 5$\sigma$ error bar for pure LAB (in dashed black), LAB + 2.5 g/L PPO (in dash-dotted green) and the JUNO mixture (in violet), measured with three different optical paths.}
\label{fig:AbsL}
\end{figure}

\noindent We then converted the absorbance of different mixtures to absorption length, by exploiting equation \ref{eq:absorbance}.
In order to reach a precision better then 5$\%$ and to neglect the effect of other unwanted phenomena beside absorption (scattering, fluorescence, dark noise, etc.) we imposed some limitations to the absorbance, 0.04$\ <A< \ $ 1.3, which implies a limitation on the absorption length as well, as studied in \cite{Wen}. In order to extend our sensitivity to a wider range of absorption lengths, we performed multiple measurements using cuvettes increasing the optical path (1, 5 and 10 cm), which is linked to the absorption length sensitivity by equation \ref{eq:absorbance}.
Figure \ref{fig:AbsL} shows the weighted mean trend for the three mixtures, pure LAB in black dashed line, LAB + 2.5 g/L PPO in green dash-dotted line and the JUNO mixture in violet solid line.

\noindent Note that given the reduced length of our setup cuvettes, our sensitivity on the absorption lengths is limited to $\leq$ 1 m. However, considering the rapid increasing trend at $\sim$ 405 nm, we expect the absorption length to be $\geq$ 18 m for wavelengths $\geq$ 430 nm, consistent with the results observed in liquid scintillators of similar composition and similar solvent \cite{Att_Length, PALM, DB_abs}. In addition, due to aging it is possible that the optical properties, in particular the attenuation length, will change over time. For this reason some experiments have been conducted, showing that in 4 years the LS performances are preserved with small effects on the attenuation length \cite{Aging}. Therefore for a long living experiment like JUNO a careful calibration of the evolution of the optical properties of the scintillator inside the detector is mandatory.

\section{Timing properties}\label{sec:fluorescence}

The timing properties can be investigated in a dedicated setup, by exploiting radioactive sources. 

\noindent The time distribution of the fluorescence light emitted by the liquid scintillator can be modelled as a superposition of different decreasing exponential distributions with characteristic time constants $\tau_i$ and relative weights $q_i$ . The fastest scintillation light component from the liquid scintillator is expected to show a time constant $\tau_1$ at the nanosecond level, while the others range from $\sim$ few nanoseconds to hundreds of nanoseconds.

\noindent We used the Time-Correlated Single Photon Counting technique to measure the characteristic times $\tau_i$ \cite{PAOLO, TCSPC, Hans}. This well-established technique requires that each event is detected by two PMTs, one strongly coupled, the other one weakly coupled to the scintillator. The first one detects the signal from the very first photons emitted, determining the start of the event, while the second one observes only one photon at a time. This random selection allows to reconstruct the entire time profile around one million events. 


\subsection{Setup}
\label{sub:Setup_timing}
\begin{figure}
\centering
\includegraphics[width=0.5\columnwidth]{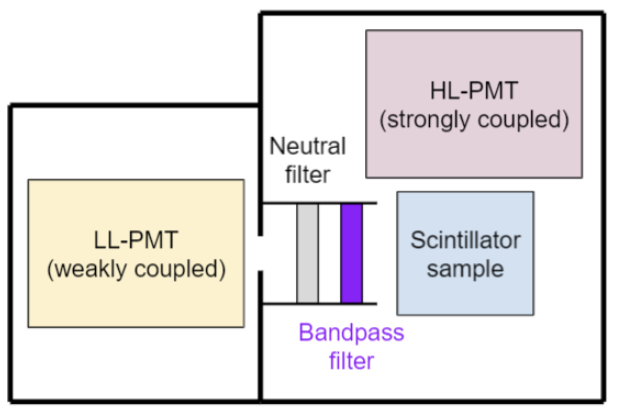}
\caption{Scheme of the experimental setup used for the fluorescence time profile measurements. The liquid scintillator sample is contained in a black box and it is viewed by two PMTs. The strongly coupled PMT, which looks directly to the sample, and the weakly coupled PMT, which looks to the sample through a neutral or a band-pass filter to reach the single photon counting regime.}
\label{fig:Scheme}
\end{figure}

The setup for the measurement of the timing features of the light emitted by the liquid scintillator is shown in figure \ref{fig:Scheme} and is basically composed by a cuvette (internal dimensions: $3.1 \times 2.8 \times 3.3 \, \mathrm{cm^3}$), made of optical glass, filled with the liquid scintillator mixture and two PMTs facing the cuvette inside a black box. One PMT (model Hamamatsu R1828-01) called High Level (HL-)PMT in figure \ref{fig:Scheme}, is part of an assembly (Hamamatsu H1949-51) including a magnetic shield. It is placed close to the cuvette, facing it, so that its optical coupling to the liquid scintillator is very strong. A second PMT (model Hamamatsu R4220P), called Low Level (LL-)PMT in figure \ref{fig:Scheme}, is a side-window PMT optimized for single-photon counting applications. This latter is placed far from the cuvette, facing it, and different filters can be positioned between the cuvette and the PMT to further reduce the optical coupling to the liquid scintillator. A fundamental hypothesis to exploit the Time-Correlated Single Photon Counting technique is that the LL-PMT must detect a photon at most in $\sim 3\%$ of cases compared to the HL-PMT.
To ensure this condition we always used a NDUV13A neutral density optical filter (optical density: 1.3). In addition, the spectrum of the charge collected by the PMTs is acquired simultaneously with the time profile measurement allowing to check the single photon counting regime in real time.

\noindent The cuvette is enclosed in an aluminum frame and an  air-tight cap closes its open face on top. The cap is provided with a needle that goes all the way into the liquid scintillator and allows nitrogen bubbling. Nitrogen bubbling has proven to be effective in reducing quenching caused by oxygen dissolved in liquid scintillators \cite{PAOLO}. This operation is repeated before each new measurement by flushing the sample for a minimum of half an hour to remove the residual oxygen.

\noindent The box containing the two PMTs and the cuvette is placed between two plastic scintillator modules, one above and one below, with the goal of vetoing cosmic-ray muons. Each module is composed by a slab of EJ-200 (500 mm $\times$ 500 mm $\times$ 20 mm) wrapped in reflective foil, enclosed in a black vinyl light-tight container and coupled to a 25 mm - diameter PMT at one corner of the scintillator.


\noindent The signals coming from the PMTs facing the liquid scintillator are digitized by two separate NI PXIe-5162 digitizers (10 bit, 5 GS/s, 1.5 GHz) included in a NI PXIe-1075 chassis controlled by a NI PXIe-8135 embedded controller.
The signals coming from the veto modules pass through a NIM coincidence unit whose output is digitized by a third NI PXIe-5162 module.
The digitizers are operated in interleaving mode to fully exploit their capabilities and to maximize the sampling rate.

\noindent The setup is also equipped with different sources, in particular we used $^{244}\mathrm{Cm}$ as an $\alpha$ source and $^{60}\mathrm{Co}$ as a $\beta$ source. Note that $^{60}\mathrm{Co}$ emits both a $\beta$ (electron) and gamma particles. The electrons are completely shielded by the capsule containing the source, while gammas can reach the scintillator, producing electrons by Compton scattering.

\subsubsection{Instrument Response Function}
\label{sub:IRF}

\begin{figure}
\centering
\includegraphics[width=0.7\columnwidth]{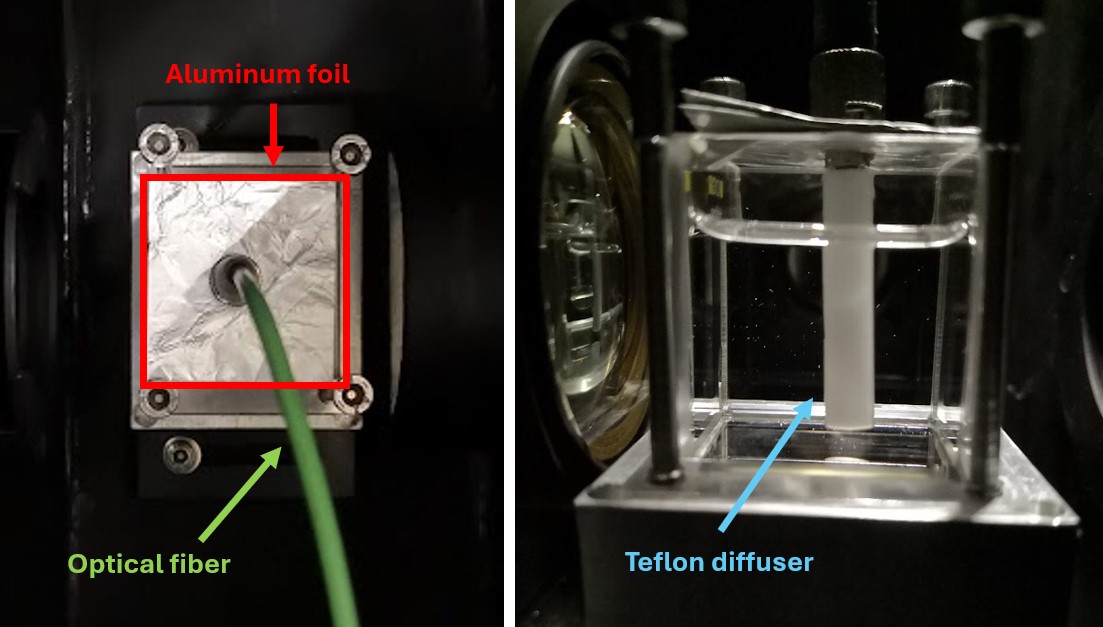}
\caption{Top (left image) and front (right image) view of the cuvette containing the pure LAB and the diffuser for the IRF measurement. The left image also shows the optical fiber and the aluminum foil used to cover the cuvette. } 
\label{fig:diffuser}
\end{figure}

The Instrument Response Function (IRF) of the entire system, that is the response to a delta function, has been measured using a pulsed laser source with 75 ps pulse width and a wavelength of 405 nm. The laser was coupled to an optical fiber entering the black box, terminated with a Teflon diffuser.
The light scattered from the diffuser, conveniently attenuated by means of an optical filter, mimics the emission from the scintillator and it is sufficiently faint to fulfill the single-photon condition in the weakly-coupled PMT, yet intense enough to cause the strongly-coupled PMT to trigger the acquisition.
To better emulate the effect of refractions and reflections, the diffuser was placed inside a cuvette containing LAB\footnote{We used pure LAB and not the full scintillating mixture, because a small fraction of the laser light could have been absorbed and re-emitted by the JUNO scintillator, spoiling the time response analysis.} and covered with an aluminum foil, so that the conditions were as close as possible to the fluorescence time measurement (figure \ref{fig:diffuser}).


\begin{figure}
\centering
\includegraphics[width=0.7\columnwidth]{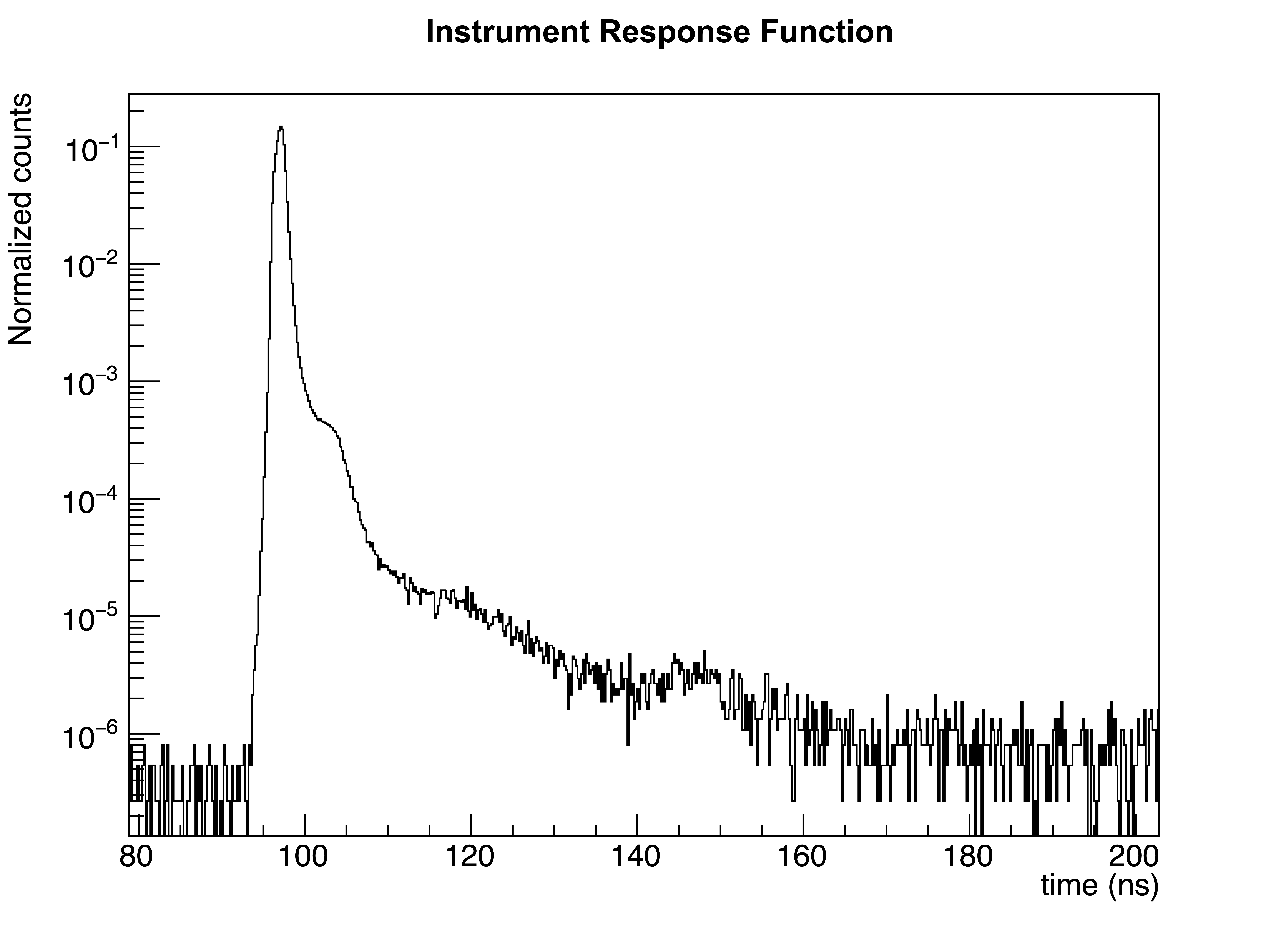}
\caption{Measurement of the Instrument Response Function of the experimental setup for time profile measurements. The FWHM of the main peak is less than $1\,\mathrm{ns}$.}
\label{fig:IRF}
\end{figure}

\noindent The IRF of our experimental setup is shown in figure \ref{fig:IRF} in logarithmic scale. The full width at half maximum of the main peak of the time distribution is less than 1 ns. All the structures in the IRF are due to the after pulses and late pulses of the LL-PMT and the reflections in the cuvette. It is important to notice that taking into account all the features of the IRF is crucial to reduce possible biases in the fluorescence time reconstruction. In particular, the afterpulse-induced bump at about 103 ns is placed in the region where the scintillation time profile is changing its slope; this feature must be considered to reach an high resolution on the fluorescence parameter.

\noindent The IRF has been fitted as a superposition of seven Gaussian distributions. Through a $\chi^2$ test, we confirmed that this is a sufficiently complex model to reproduce well the IRF structure. By using a toy Monte-Carlo we also verified that the seven Gaussian model ensures a description of the fluorescence time distributions with an accuracy on the order of $\sim\%$. 

\noindent We measured the IRF by placing the diffuser in three different positions inside the cuvette to study the different effect of refraction and reflections, accounting for possible systematic effects due to the geometry of our system. We found that placing the diffuser in the center of the cell is equivalent to averaging different measurements in different positions. Hence we decided to use the central position to measure the IRF shown in figure \ref{fig:IRF}.

\begin{figure}[h]
\centering
\includegraphics[width=0.5\columnwidth]{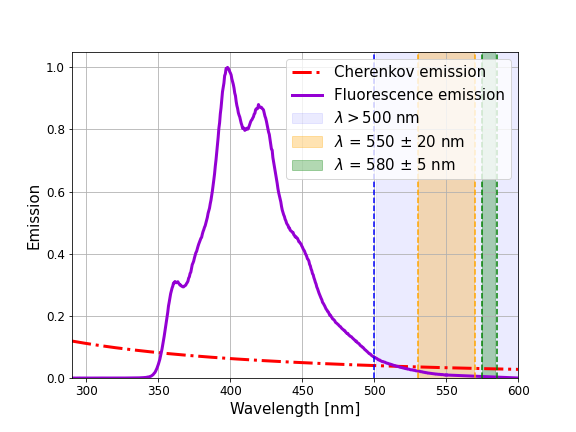}
\includegraphics[width=0.48\columnwidth]{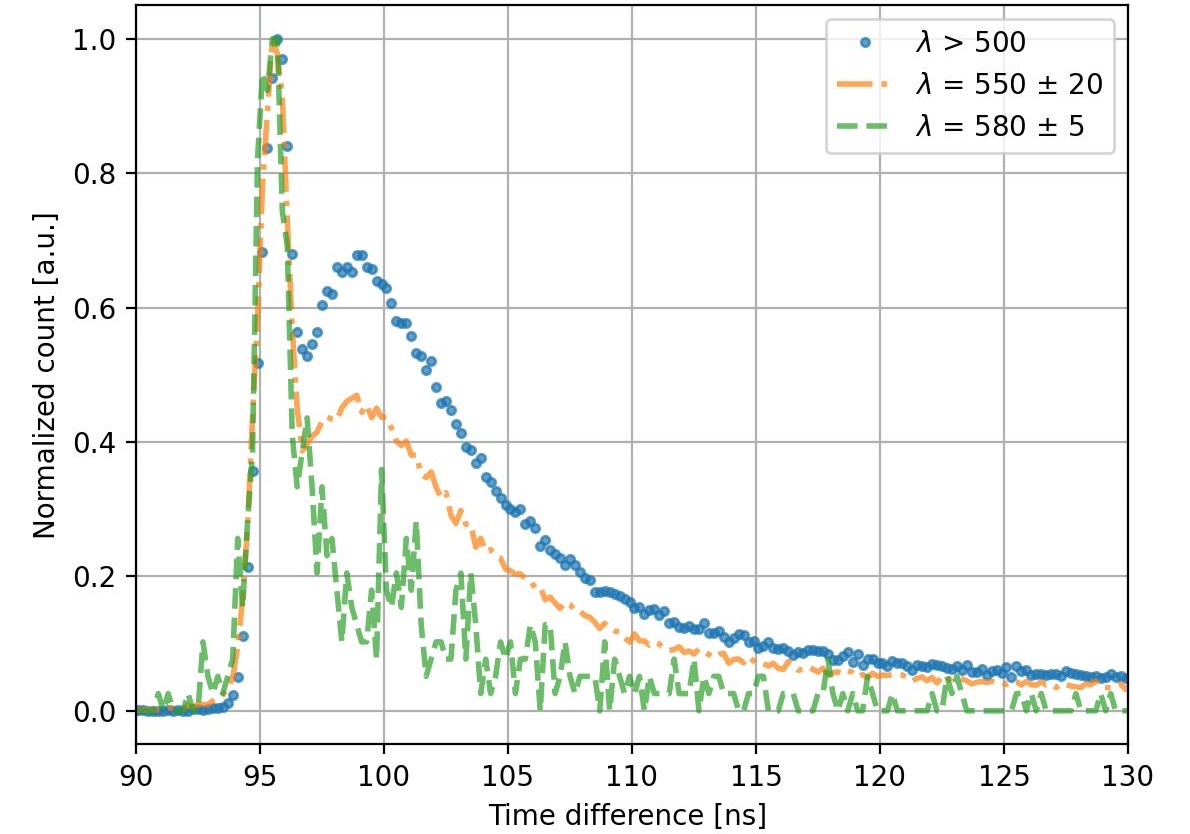}
\caption{The left plot shows the filter ranges with respect to the liquid scintillator emission spectrum normalized to the maximum (violet solid line), and the Cherenkov (red solid line, not in scale) emission spectrum. The right plot shows the comparison  between the Cherenkov measurements obtained with  500 nm long pass filter (in dotted blue line), 550 nm  $\pm$ 20 nm (in dashed-dotted orange line) and 580 nm  $\pm$ 5 nm (in dashed green line) band-pass filters. } 
\label{fig:Cherenkov}
\end{figure}

\noindent In order to have a crosscheck of our IRF measurement we also exploited the Cherenkov emission in JUNO liquid scintillator.
The Cherenkov light is emitted when an ionizing particle travels faster than the speed of light in a medium and it causes the emission of photons in a $\sim$ ps time scale, which is practically negligible for our experimental setup and therefore can be considered as emitted instantaneously as the particle moves.
While the scintillator emission spectrum is concentrated in a limited wavelength range (see the left plot of figure \ref{fig:Cherenkov}), the Cherenkov emission spectrum extends to higher wavelengths. Therefore, if we select high wavelengths ($\lambda >$530 nm for JUNO liquid scintillator), it is possible to separate the Cherenkov contribution from fluorescence. 

\noindent In order to perform the Cherenkov light measurements we exploited $\isotope[60][]{Co}$ sources, placed outside the cell, and the NDUV13A filter was substituted with different filters: 550 nm  $\pm$ 20 nm  and 580 nm  $\pm$ 5 nm band-pass filters and a 500 nm long pass filter.
The right plot of figure \ref{fig:Cherenkov} shows the three measurement performed. In particular we can observe, as expected, that the Cherenkov contribution increases with respect to the fluorescence one as the wavelength range of the filter increases.
Even if this type of measurement does not allow to recreate the detailed structure of the IRF, by fitting the Cherenkov peak with a gaussian function we demonstrate that it does not change with the wavelength. Indeed, for all the wavelength ranges considered, its FWHM is comparable with the IRF one measured with the laser ($\sim$ 1 ns), further consolidating our results. 

\subsection{Fluorescence time measurements} 
We are interested in measuring the fluorescence induced in the JUNO liquid scintillator by $\alpha$ and $\beta$ radiations. To evaluate the $\alpha$ induced time profile we used a $^{244}\mathrm{Cm}$ radioactive source screwed to the cell cap and immersed inside the liquid scintillator. This allows us to be sure that the $\alpha$-particle can reach the liquid scintillator to ionize it. Instead, for what concerns the $\beta$ radiation we used a $^{60}\mathrm{Co}$ source placed outside the cell. 
To perform our measurements, we used the signal induced by the strongly-coupled PMT to give the start time of the event. Hence a DAQ window of 1600 ns was opened and we waited a signal from the single-photon PMT to stop the acquisition. This particular time window was specifically chosen to include a significant part of the time profile tails. 
\begin{figure}[h]
\centering
\includegraphics[width=0.45\columnwidth]{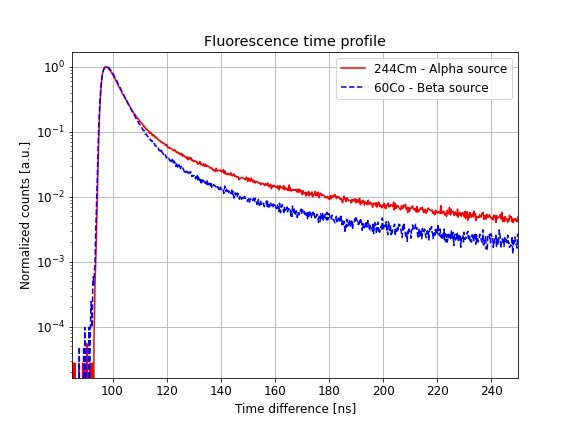}
\includegraphics[width=0.45\columnwidth]{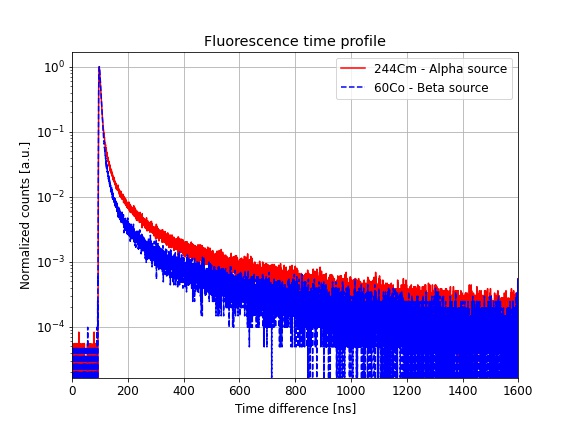}
\caption{Fluorescence time profile of the JUNO liquid scintillator induced by an $\alpha$ source (in red solid line) and a $\beta$ source (in blue dashed line). The profiles are normalized at the maximum to enhance the differences between alpha and beta profiles. Note that the main peak of the profiles is set at about $\sim 95$ ns by our acquisition software in order to observe possible effects that precede the peak.}
\label{fig:TimeProfiles}
\end{figure}

\noindent The fluorescence time profiles of the JUNO liquid scintillator, measured with our setup, are shown in figure \ref{fig:TimeProfiles}. As expected, the particles with the highest $dE/dx$ (energy loss per path length unit), like alphas, have a longer tail with respect to particles with low $dE/dx$, like electrons. This opens the possibility to perform particle identification via Pulse Shape Discrimination (PSD) techniques \cite{PAOLO}, as it will be outlined in section \ref{subsec:psd}.


\subsection{Data analysis}
The time distribution of the fluorescence light emitted by the scintillator can be described as a linear combination of decreasing exponential distributions.
We started by modelling the time distribution of emitted light with four effective components, as it was already done in literature for LAB-based scintillators \cite{PAOLO}. We checked that this is the best choice (see the end of this paragraph).
Thanks to the high time resolution of our setup, we have been able to observe the rising edge of the light signal and this had to be taken into account in our analysis.
The resulting distribution can be written as:
\begin{equation}
F_{fit}(t) = N \sum_{i=1}^4\frac{q_i}{\tau_i-\tau_r}\left(e^{-\frac{t-t_0}{\tau_i}}- e^{-\frac{t-t_0}{\tau_r}} \right) \Theta(t-t_0)
\label{Eq: FitModel}
\end{equation}
where $N$ is a normalization constant, $q_i$ and $\tau_i$ are the relative weights (also called ``fractions" hereafter) of the $i$-th component and its characteristic decay times, $\tau_r$ is the characteristic rise time (common to all the four components) and $\Theta(t-t_0)$ is the step function that marks the start of the fluorescence time profile.
The normalization of the fourth component requires a dedicated treatment due to the acquisition time window, which is limited to 1600 ns. This limitation implies a truncation in the light collection, so the normalization of the fourth exponential component $\frac{q_4}{\tau_4-\tau_r}$  needs to be corrected:
\begin{equation}
\tilde{ q_{4}}  = \frac{1-\sum_{i=1}^3q_i}{1-e^{-\frac{t_w-t_0}{\tau_4}}} 
\label{Eq: q4}
\end{equation}

\noindent where $t_w$ = 1600 ns is the length of the DAQ time window.
The fraction of light uncollected because of the finite duration of our acquisition window is about $9.1\%$ for the fourth component, while it is negligible for the other three contributions.

\noindent To take into account the finite resolution of the setup it is necessary to convolve the fluorescence model described in equation \ref{Eq: FitModel} with the IRF shown in figure \ref{fig:IRF}: 
\begin{equation}
(F_{fit}*\mathrm{IRF})(t) = \int_{-\infty}^{\infty}F_{fit}(\tau)\mathrm{IRF}(t-\tau)d\tau
\end{equation}
Thanks to the linearity of the operation, the analytic convolution of the IRF with the fluorescence model in equation \ref{Eq: FitModel} gives:
\begin{equation}
F_{fluo}(t)=N\sum_{i=1}^4\sum_{j=1}^7N_iN_j\left(e^{-t/\tau_i}-e^{-t/\tau_r}\right)* G_j(t; \mu_j, \sigma_j)
\label{eq:F_fluo}
\end{equation}
where the $N_i$ is the normalization of the i-th fluorescence component, as discussed before, $N_j$ are the normalized relative weights of the seven Gaussian functions and $N$ is the total number of entries in the histogram. 

\begin{figure}[h]
\centering
\includegraphics[width=0.435\columnwidth]{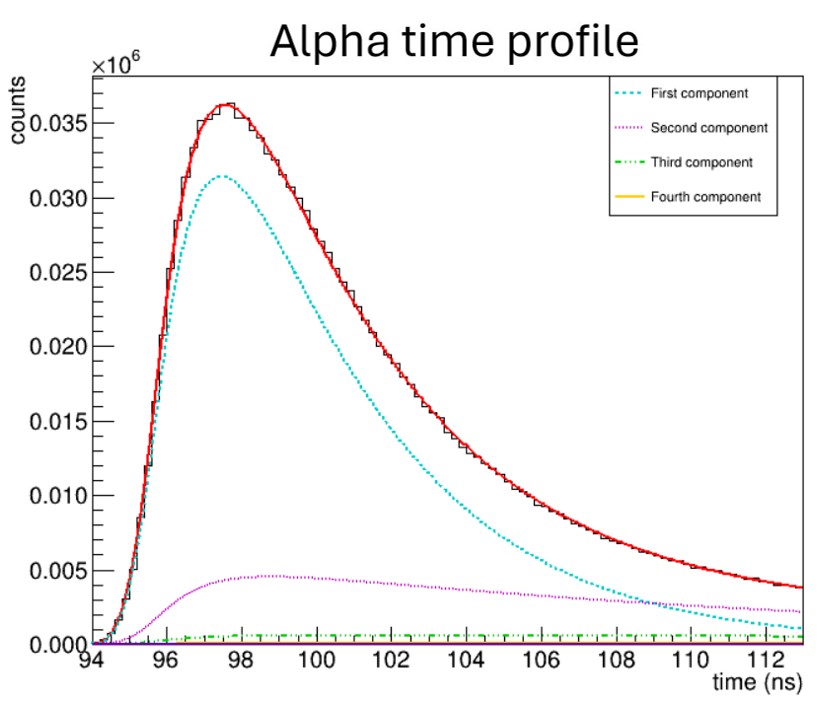}
\includegraphics[width=0.52\columnwidth]{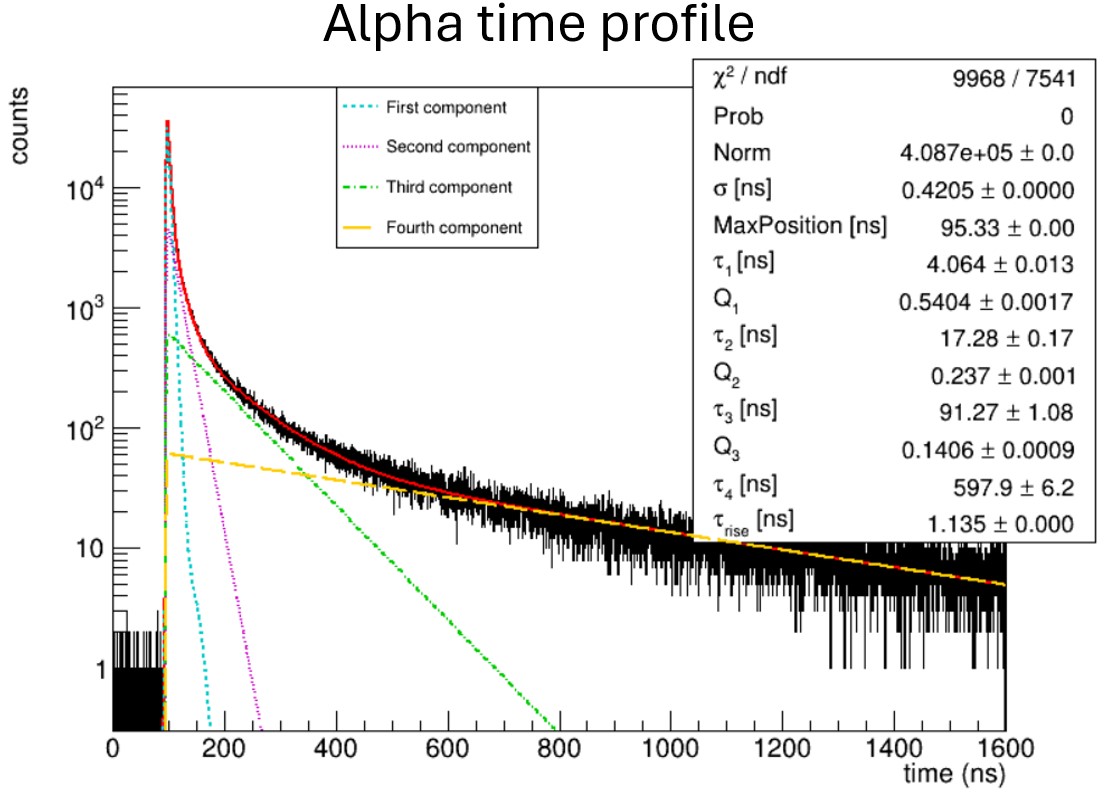}
\caption{Example of a fit on data measured using a $^{244}\mathrm{Cm}$ source. The data are shown in black. The total fit function is shown in red solid line, while the first, the second, the third and the fourth components are shown in cyan dashed line, magenta dotted line, green dash-dotted line and yellow wide-dashed line respectively. The left plot shows a zoom of the peak region dominated by the first and second components.}
\label{fig:AlphaFit}
\end{figure}

\noindent An example of the fit superimposed on the data from an $\alpha$ measurement is shown in figure \ref{fig:AlphaFit}.

\noindent In order to evaluate the validity of the four exponential model we also tried to fit our data assuming a model based on three and five exponentials. In the former case we observed a worse $\chi^2$, while in the latter one the fifth component contribution railed to zero. Furthermore,it is important to notice that the IRF parameters were fixed during the fitting procedure of the fluorescence curve to the value obtained by analyzing the IRF curve.

\noindent Our results for the fluorescence parameters of the JUNO scintillator are collected in table \ref{tab:Resutls}. It is possible to notice that, as expected, the higher is the $dE/dx$ of the incident particle, the higher are the characteristic decay times $\tau_i$, but also the $q_3$ and $q_4$ values (both corresponding to the slow fluorescence component), resulting in a higher tail.


\begin{table}[h!]
\caption{Best estimation of the fluorescence time parameters for $\alpha$ and $\beta$ profiles}
\label{tab:Resutls}
\centering
\resizebox{0.8\columnwidth}{!}{
	\begin{tabular}{l|c|c|c|c|}
		\cline{2-5}
		& \textbf{$\tau_1$ {[}ns{]}} & \textbf{$\tau_2$ {[}ns{]}} & \textbf{$\tau_3$ {[}ns{]}} & \textbf{$\tau_4$ {[}ns{]}} \\ \hline
		\multicolumn{1}{l|}{$\alpha$} & 4.06\,±\,0.01                  & 17.29\,±\,0.17                 & 91.3\,±\,1.1                   & 598\,±\,6                     \\ \hline
		\multicolumn{1}{l|}{$e^-$}    & 3.86\,±\,0.02                  & 14.52\,±\,0.25                 & 81.3\,±\,1.8                   & 570\,±\,10                      \\ \hline
		& \textbf{$q_1$ {[}\%{]}}    & \textbf{$q_2$ {[}\%{]}}    & \textbf{$q_3$ {[}\%{]}}    & \textbf{$q_4$ {[}\%{]}}    \\ \hline
		\multicolumn{1}{l|}{$\alpha$} & 54.04\,±\,0.17                 & 23.70\,±\,0.14                 & 14.06\,±\,0.09                 & 8.92\,±\,0.26                  \\ \hline
		\multicolumn{1}{l|}{$e^-$}    & 67.79\,±\,0.35                 & 20.01\,±\,0.27                 & 7.53\,±\,0.11                  & 5.03\,±\,0.50
		\\ \hline
	\end{tabular}%
}
\end{table}

\noindent In order to test the reproducibility of our measurement we repeated it several times using the same mixture and we obtained compatible results within errors.  

\noindent As we highlighted in section \ref{sub:IRF}, in addition to fluorescence also Cherenkov light is emitted.
The Cherenkov fraction measured with filters at 500 nm, 550 nm and 580 nm are $\sim9\%$, $15\%$ and $>25\%$ respectively (see figure \ref{fig:Cherenkov}, right plot). From these measurements we estimate the Cherenkov fraction present in the region dominated by fluorescence (from 350 nm to 550 nm) to be $< 1\%$. Using a toy Monte-Carlo we verified that, for our setup, this fraction does not affect the fluorescence time profiles. 

\subsection{Pulse Shape Discrimination performances}\label{subsec:psd}

\begin{figure}[h]
\centering
\includegraphics[width=0.55\columnwidth]{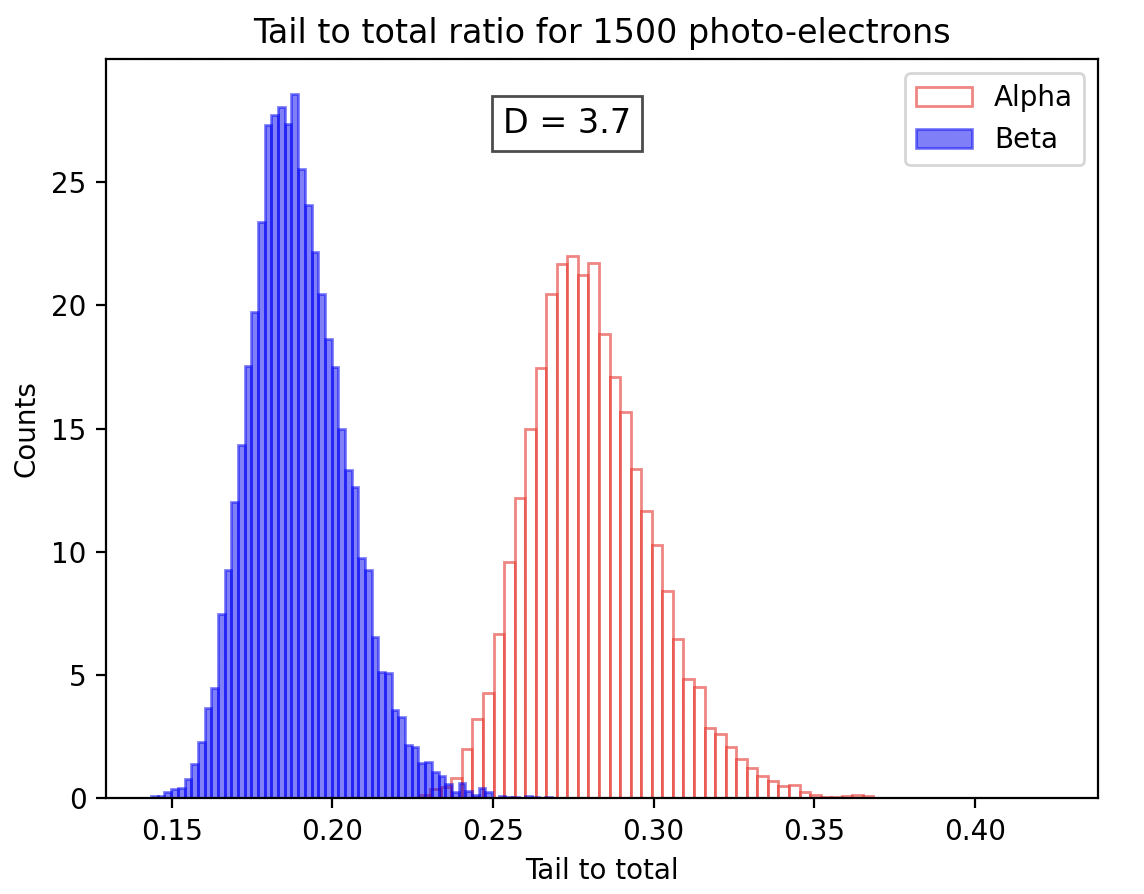}
\caption{Example of tail to total distribution for both $\alpha$ (red) and $\beta$ (blue) radiations, assuming 1500 photo-electrons (which correspond to a $\sim1$ MeV event in JUNO). The $t_{cut}=70\,\mathrm{ns}$ was optimized to have the maximal separation between the two distributions}
\label{fig:TtoT}
\end{figure}

The fluorescence time profiles for alphas and betas shown in figure \ref{fig:TimeProfiles} are significantly different: in particular, the scintillation light induced by alpha particles has proportionally more photons emitted at longer time with respect to the one induced by beta particles.
This opens the possibility to perform particle identification via the so-called Pulse Shape Discrimination (PSD) technique \cite{PAOLO}, which plays an important role for background rejection. For example, the detection of solar neutrinos via elastic scattering on electrons  (which provides a beta-like signal) can greatly benefit from the rejection of alpha radioactive background.
We have estimated the alpha/beta discrimination capability  of the JUNO liquid scintillator by using a simple algorithm, based on the so-called ``tail-to-total" ratio\footnote{Note that  the ``tail-to-total" method is the simplest PSD technique. Better results could be achieved using more sophisticated techniques, like the Gatti or Neural Network based  methods \cite{BX_PSD}} $R$. $R$ is calculated for each scintillation event as the number of photons collected in the tail of the time distribution (for $t>t_{cut}$) over the total amount of photons: in average, $R$ is larger for alpha events with respect to beta events. We used a toy Monte-Carlo to simulate $10^4$ of alpha and beta events, following the measured profiles of figure \ref{fig:TimeProfiles}, and we built the distributions of $R$. The $R$ distributions obtained with this procedure for beta and alpha events (blue dashed and red solid line respectively) are shown in figure \ref{fig:TtoT}. In this example, we assumed 1500 collected photon-electrons for each event corresponding to about 1 MeV of deposited energy in JUNO. 

\noindent To quantify the separation between the two distributions, and therefore the alpha/beta discrimination capability, we define the merit factor $D$ as follow:
\begin{equation}
D = \frac{\mu_{\alpha} - \mu_{\beta}}{\sqrt{\sigma_{\alpha}^2 + \sigma_{\beta}^2}}
\label{Eq: D}
\end{equation}
where $\mu_{\alpha}$ and $\sigma_{\alpha}^2$ represent the mean value and the variance of the $\alpha$ distribution respectively, likewise for electrons.

\noindent The larger the value of $D$, the greater the separation between the two curves, and therefore the alpha/beta discrimination capability. In the case shown in figure \ref{fig:TtoT}, the factor of merit is D=3.7, which would for example  allow to reject 99.5\% of alphas (background) while retaining more than 99\% of the betas (signal). 

\begin{figure}[h]
\centering
\includegraphics[width=0.5\columnwidth]{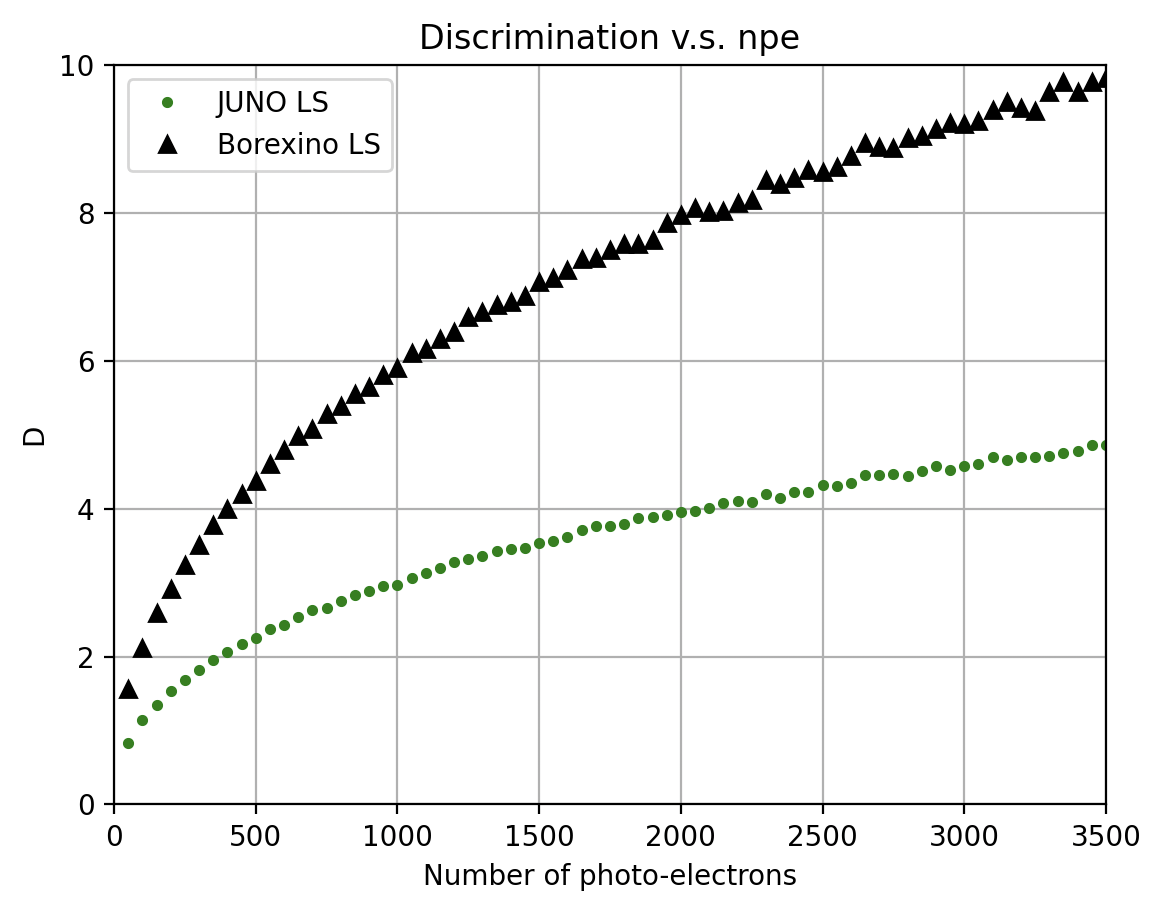}
\caption{Factor of merit $D$ as function of the number of photo-electrons produced for both the JUNO (green dots) and the Borexino (black triangles) liquid scintillators.}
\label{fig:psd_performances}
\end{figure}

\noindent The discrimination capability increases with the number of collected photons as can be seen in figure \ref{fig:psd_performances}.

\noindent Figure \ref{fig:psd_performances} shows also a comparison between the JUNO liquid scintillator and the Borexino scintillator (pseudocumene + 1.5 g/l of PPO), measured with the same setup \cite{PAOLO}.
The Borexino mixture performs better than the JUNO one. However, the number of photo-electrons detected by JUNO for a given energy will be a factor $\sim3$ larger with respect to Borexino, mostly because of the larger optical coverage and the higher Q.E. of the phototubes. This makes the alpha/beta discrimination performance of the two detectors comparable for similar energies. 
Finally, it is important to underline that the results of figure \ref{fig:TtoT} and \ref{fig:psd_performances} are made on small  scintillator  samples  and therefore do not take into account light propagation effects (absorption with subsequent re-emission, scattering) which could worsen significantly the performance of the PSD method in a large detector like JUNO. However, in order to properly include these effects, the full JUNO Monte-Carlo simulation would be needed, which is beyond the scope of this article.

\section{Conclusions}\label{sec:conclusions}
The aim of this work is to provide a precise and accurate measurement of the fluorescence emission of the JUNO liquid scintillator (LAB + 2.5 g/L PPO + 3 mg/L Bis-MSB). First of all we studied both the emission and absorption of the JUNO liquid scintillator mixture, by measuring its emission spectrum and its absorption length. We compared them with those of LAB and LAB + 2.5 g/L PPO to highlight how the PPO and the Bis-MSB shift the emission and the absorption of the LAB to higher wavelengths. In particular we verified that the emission spectrum of the JUNO liquid scintillator matches the quantum efficiency of JUNO large NNVT-PMTs.
We then used a custom made setup to measure the $\alpha$ and $\beta$ time profiles for JUNO liquid scintillator, which will be fundamental for the $\alpha/\beta$ discrimination in JUNO. The detector response has been accurately measured with laser light and crosschecked exploiting the Cherenkov light. By adopting a four exponential model we measured the characteristic time constants $\tau_i$ and relative weights $q_i$ both for $\alpha$ and $\beta$ profiles. In particular, the fluorescence parameters have been obtained with an accuracy of $\sim 1\%$ for the fastest components and of few percent for the slowest ones. 
We have also checked that the Cherenkov contribution does not affect the fluorescence time measurements.
We then estimated the Pulse Shape Discrimination performances of the JUNO liquid scintillator, based on a small scale analysis, with the `tail to total' ratio method. The results are promising and showed the possibility to reject 99.5\% of alphas background while retaining more than 99\% of the betas (signal).\\
Our results will be fundamental to build a reliable Monte-Carlo code capable of reproducing the response of the JUNO detector. This is essential to ensure the optimal energy and position reconstruction required for the Neutrino Mass Ordering determination in JUNO.

\section*{Acknowledgments}
This work was realized thanks to the University of Milan (UNIMI) and the National Institute of Nuclear Physics (INFN), Milan section. D. Basilico, M. Grassi, and A.C. Re  acknowledge that this study was partially funded by MUR PRIN 2022 “2022NAE9AJ - Getting ready to capture an exploding star” (Italy). In addition, we thank  the JUNO liquid scintillator working group for useful discussions and practical help in producing the liquid scintillator, in particular, Hans Steiger, Raphael Stoch, Meishu Lu and the IHEP group.


\end{document}